\documentclass[]{interact}

\usepackage{geometry}
\usepackage[labelsep=period]{caption}

\usepackage{array}
\usepackage{booktabs}
\usepackage{bm}
\usepackage{amsthm}
\usepackage{amsmath}
\usepackage{amstext}
\usepackage{graphicx}
\usepackage[natbibapa,nodoi]{apacite}
\usepackage{hyperref}
\usepackage{lineno}
\usepackage{bbm}
\usepackage{url}

\usepackage{algorithm}
\usepackage{algorithmicx}
\usepackage{algpseudocode}
\usepackage{verbatim}

\usepackage{setspace}
\usepackage{breakurl}

\usepackage{hyperref}
\hypersetup{
  colorlinks=true,
  linkcolor=blue,
  citecolor=blue,
  urlcolor=blue}

\usepackage{bm}

\usepackage{enumitem}
\begin{document}

\title{The uncertainty estimation of feature-based forecast combinations}

\author{
\name{Xiaoqian Wang \textsuperscript{a}, Yanfei Kang \textsuperscript{a}, Fotios Petropoulos \textsuperscript{b} and Feng Li \textsuperscript{c} \thanks{CONTACT Feng Li. Email: feng.li@cufe.edu.cn}}
\affil{\textsuperscript{a}School of Economics and Management, Beihang University, Beijing, China; \textsuperscript{b}School of Management, University of Bath, UK; \textsuperscript{c}School of Statistics and Mathematics, Central University of Finance and Economics, Beijing, China.}
}

\maketitle

\begin{abstract}
Forecasting is an indispensable element of operational research (OR) and an important aid to planning. The accurate estimation of the forecast uncertainty facilitates several operations management activities, predominantly in supporting decisions in inventory and supply chain management and effectively setting safety stocks. In this paper, we introduce a feature-based framework, which links the relationship between time series features and the interval forecasting performance into providing reliable interval forecasts. We propose an optimal threshold ratio searching algorithm and a new weight determination mechanism for selecting an appropriate subset of models and assigning combination weights for each time series tailored to the observed features. We evaluate our approach using a large set of time series from the M4 competition. Our experiments show that our approach significantly outperforms a wide range of benchmark models, both in terms of point forecasts as well as prediction intervals.
\end{abstract}


\newpage
\setcounter{page}{1}

\section{Introduction}
\label{sec:intro}
With the advent of the big data era, a large amount of time series data is being continuously collected, which has led to an explosive demand for time series forecasting methods. Time series forecasting has played a pivotal role in the development of many fields, such as finance, meteorology, and signal processing. The vast majority of the time series forecasting literature aims to improve point forecasting accuracy, and they mainly forecast the mean or the median of the distributions for future observations. However, more attention should be focused on quantifying the uncertainty of the prediction to measure the reliability of the forecasting results. As a result, there is a large demand in many fields of research for forecasting methods that can provide a comprehensive outlook of the expected future values and the future uncertainty.

Forecasting is an indispensable element of operational research (OR) \citep{Fildes2008}. In a recent article, \cite{Nikolopoulos2020-zi} mentions that ``we have no other option rather than throwing as many examples as possible of how OR changes our lives [...] within the ubiquitous OR discipline, forecasting is the finest example.'' He continues to enlist a series of application areas of forecasting in OR, such as healthcare, tourism, and marketing. Within OR, applications of estimating forecast uncertainty include finance \citep{Tung2009}, energy \citep{Taylor2017-sk}, supply chains \citep{Rahman2011}, and inventory management \citep{inventoryreview}.

As claimed by the \emph{no-free-lunch} theorem \citep{wolpert1997no}, it is impossible for all forecasting methods to perform well on all time series. \citet{petropoulos2014horses} also point out that there are \emph{horses for courses}, and appropriate forecasting methods have to be chosen for certain time series. \citet{cang2014combination} argue that forecast combination is superior in forecasting accuracy to the individuals used for averaging. \cite{Petropoulos2018-cl} show that handling forecast uncertainty entails the understanding of its three sources: model uncertainty (selecting the correct model), parameters uncertainty (correctly estimating the values of the model's parameters), and data uncertainty (inherent noise/unpredictable component of time series). They find that solely tackling model uncertainty leads to significant performance improvements, giving support on the value of forecast combinations.

A vast majority of literature uses features to capture time series characteristics. A time series feature can be any statistical representation of time series characteristics (e.g., mean, standard deviation, autocorrelation and seasonality). Feature-based time series representations have received emerging interests in various time series mining tasks~\citep{kang2017visualising}, such as time series clustering, classification, and anomalous detection. Another remarkable application of features in time series analysis is feature-based time series forecasting, which focuses on associating the time series features with forecasts and utilizing this connection to improve point forecasting accuracy. \citet{petropoulos2014horses} identify the decisive effect of seven time series features on forecasting accuracies of several methods and translate these findings into forecasting method selection. \citet{talagala2018meta} propose the FFORMS (Feature-based FORecast-Model Selection) framework that identifies the best forecasting model by using time series features based on a random forest. \citet{montero2020fforma} conduct a model combining process based on meta-learning by training weights for various individual forecasting methods according to time series features.

However, compared to point forecasting, the literature on the uncertainty estimation of feature-based time series forecasts is very limited. The M4 forecasting competition \citep{makridakis2020m4} encouraged participants to provide point forecasts as well as prediction intervals (PIs). Among the submissions, \citet{montero2020fforma} compute the point forecasts using FFORMA (Feature-based FORecast Model Averaging) and obtain the PIs by using a simple equally weighted combination of the 95\% bounds of na\"{i}ve, theta and seasonal na\"{i}ve methods. This approach ranks second in the M4 competition but does not take any time series characteristic into account when calculating the interval forecasts.

The main aim of this paper is to explore how time series features affect the uncertainty estimation of forecasts, which is measured by PIs, for various forecasting methods, and to translate these findings into an attempt to improve the performance of these PIs. To accomplish this, we use generalized additive models \citep[GAMs:][]{hastie1990generalized}, which are characterized by interpretability, flexibility, and the reduction of overfitting.
GAMs are applied to depict the relationship between time series features and interval forecasting accuracies, making interval forecasts interpretable for time series features. However, how to translate these relationship findings into the improvement of time series interval forecasting remains an important question.

In this paper, we propose a general feature-based time series interval forecasting framework for the situation where the interest lies in forecasting a set of time series and evaluating their forecast uncertainty. By adapting the scoring rule to the evaluation of interval forecasting performance, the relationship between features and interval scores is established by GAMs to obtain the optimal weights for interval forecast combination per time series. Then, point forecasts as well as PIs can be obtained by the weighted combination of forecasts calculated from a pool of individual forecasting methods. The main contributions of the paper are as follows:

\begin{enumerate}
\def\labelenumi{\arabic{enumi}.}

\item Unlike previous literature on feature-based forecasting that focuses on point forecast, our proposed approach focuses on prediction intervals, making it tightly connected with OR decision making.  Also, we depict the relationship between time series features and interval forecasting accuracies, which makes our proposed framework interpretable.

\item Taking full advantage of the relationship between time series features and interval forecasting performances, we propose a new weight estimation mechanism to assign the optimal combination weights to the individual forecasting methods for each time series. To the best of our knowledge, this is the first time that time series features are taken into account for forecast uncertainty estimation.

\item Rather than combining all the individual models in the traditional forecasting combination approach, we propose an optimal threshold ratio searching algorithm, through which we select an optimal subset of the available individual methods per time series for model combination. Experiments on the M4 competition data show that the weighted combination of individual models that are selected by the optimal threshold significantly outperforms the weighted combination of all the individual methods.

\item Our proposed approach outperforms a variety of standard forecasting benchmark methods with distinctions for both point forecasts and predictive intervals. Our approach also ranks competitively against the top submissions of the M4 competition, even if direct comparisons should be treated with care as we have had access to the test data of the competition.
\end{enumerate}

The rest of the paper is organized as follows. Section \ref{sec:method} introduces the M4 competition data that is used as the test data in the paper, and presents the general feature-based time series forecasting framework proposed for the forecast uncertainty estimation. We elaborate on the components and details of this framework towards deriving the forecast combination in Section \ref{sec:fuma}. Section \ref{sec:application} applies the proposed framework to the M4 competition data and reports the experiment results. Section \ref{sec:conclusion} concludes the paper.

\section{General framework}
\label{sec:method}

\subsection{M4 competition data}
\label{sec:m4data}

To better present the proposed forecasting framework, we first introduce our test data and use it to demonstrate each aspect of the proposed method in the following sections.

We consider the yearly, quarterly and monthly subsets of the M4 competition data as our test dataset. The recent M4 forecasting competition \citep{makridakis2020m4} is a continuation of the previous M competitions, which are a series of competitions that are devoted to identifying methods with superior forecasting performance and being inspired from the submissions to advance the forecasting theory and practice. M4 competition introduces the challenge of forecasting $100,000$ time series with different periods.

\begin{figure}
  \centering \includegraphics[width=1\textwidth]{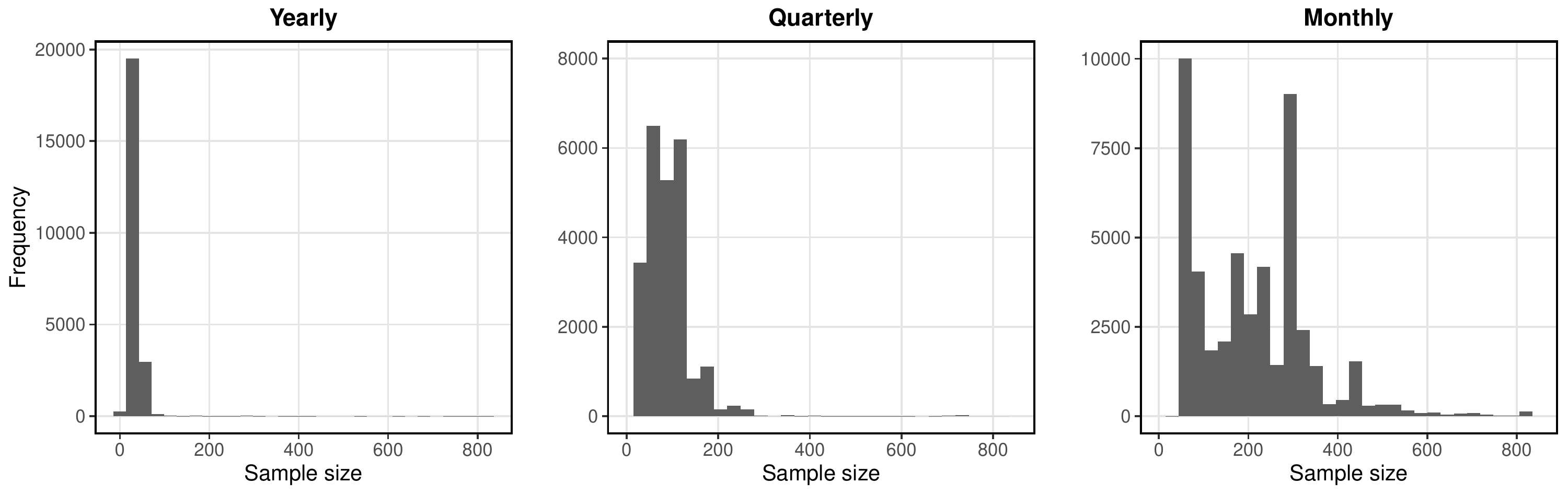}
  \caption{Distributions of sample sizes of the yearly, quarterly and monthly subsets in the M4 dataset.}
  \label{fig:hist}
\end{figure}

The M4 dataset is publicly available in the \texttt{M4comp2018} R package \citep{montero2018m4comp2018}. We focus on the yearly, quarterly, and monthly series which represent 95\% of the competition's series. The yearly subset includes $23,000$ series with sample sizes ranging from $13$ to $835$ observations and with forecast horizons of $6$ periods. The quarterly subset consists of $24,000$ series with $8$ forecast horizons, and the sample size ranges from $16$ to $866$ periods. The monthly subset contains $48,000$ time series with a constant horizon of $18$ periods ranging from $42$ to $2,794$ sample observations. As shown in Figure~\ref{fig:hist}, the sample sizes of the yearly, quarterly and monthly data in the M4 competition are not constant but vary following different distributions.

\subsection{Framework overview}
\label{sec:framework}
We propose a general feature-based time series forecasting framework for the situation where the interest lies in forecasting large collections of time series. The framework is designed mainly for providing improved uncertainty estimation of feature-based time series forecasts, which is measured by PIs. By changing the scoring rule to suit the evaluation of interval forecasting performance, we capture the relationship between time series features and the interval forecasting performance, and thus produce the weights for combining the interval forecasts from a pool of methods.

\begin{figure}
  \centering \includegraphics[width=1\textwidth]{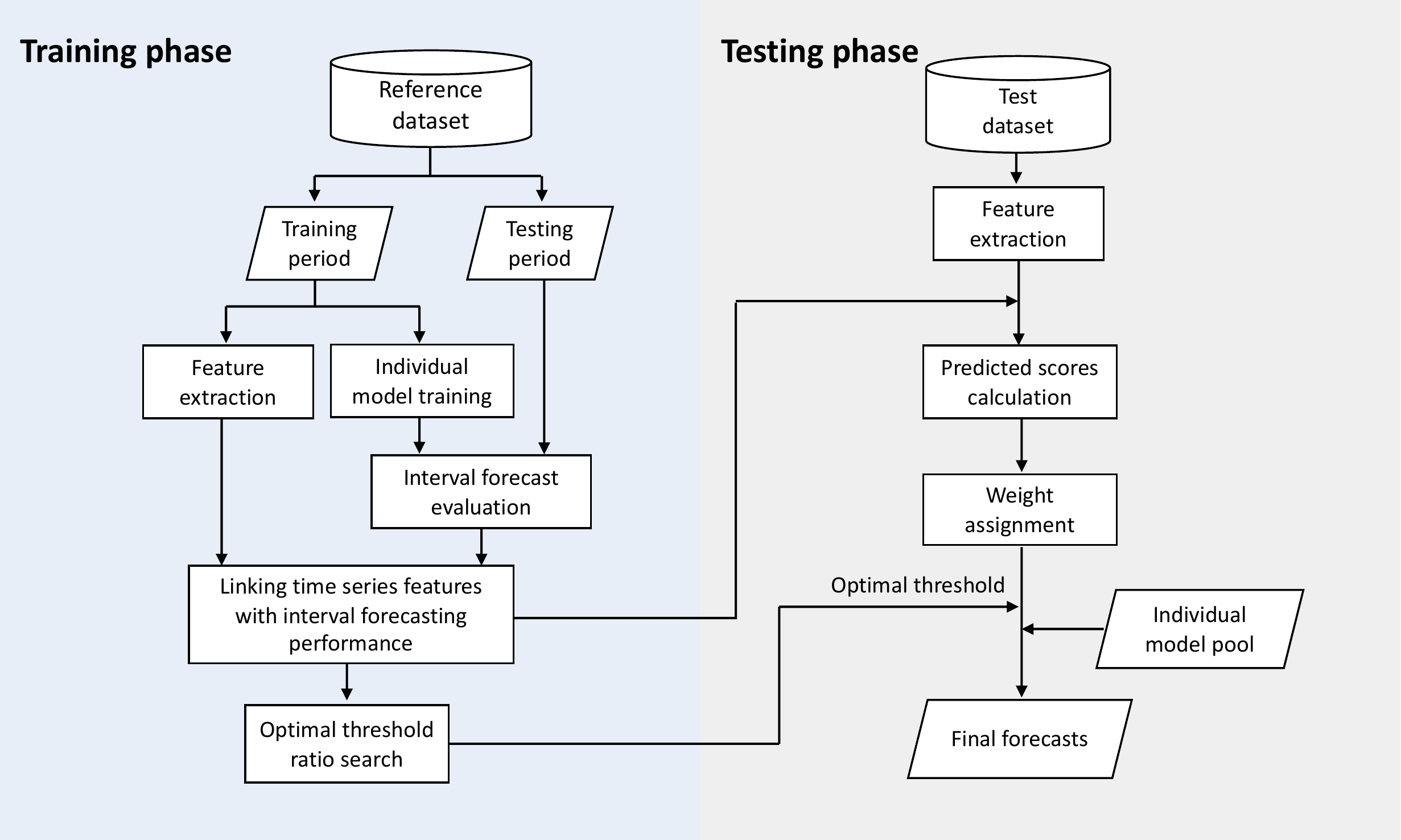}
  \caption{Feature-based time series forecasting framework. This framework is divided into training and testing phases shaded in blue and grey, respectively.}
  \label{fig:flowchart}
\end{figure}

The proposed framework, as outlined in Figure~\ref{fig:flowchart}, consists of the training and testing phases. In the training phase, a diverse set of reference data is required to train the relationship between time series features and forecasting performance of the individual methods in a pool. We describe in detail the generation of reference data in Section~\ref{sec:reference}. Given the reference dataset, we first separate each time series into a training period (historical observations) and a testing period (true values of future data). Second, we select a collection of forecasting methods (e.g., Na\"{i}ve, ARIMA) as the method pool. The training period is applied to train the individual methods and calculate the corresponding PIs. The interval forecasting performance can be easily evaluated by a certain scoring rule given the testing period. Third, we extract features (e.g., length, trend and seasonality) that reveal the intrinsic nature of the time series from the training period. Finally, we establish a model for each individual forecasting method to link time series features with its interval forecasting performance. Furthermore, we design an algorithm for the optimal threshold ratio search (see Section \ref{sec:threshold} for the details), which contributes to the identification of the advantageous individual methods used for forecast model combining.

In the testing phase, we only have to extract time series features from the newly given time series (test dataset, which is M4 in this paper) and put them into the models that have been previously trained for all the individual methods in the training phase. Subsequently, the predicted values obtained from the pre-trained models are transformed into a convex set of combining weights. The weights are then applied for model combination of the methods selected by the optimal threshold ratios. Hence, the point forecasts as well as PIs of the newly given time series are finally obtained.

It is worth mentioning that our proposed framework is a general procedure. The time series features and the forecasting method pool can be customized for the newly given time series. Moreover, we can consciously opt for a targeted approach for the time series being analyzed to capture how the time series features relate to the interval forecasting performance in our proposed framework. In this paper, we apply GAMs to achieve this goal and describe the details for our framework in the following sections.

\subsection{Reference dataset generation}
\label{sec:reference}
As shown in Figure~\ref{fig:flowchart}, our proposed framework first requires a reference dataset for the training phase. The effectiveness of our proposed framework rests on a fundamental assumption that the reference dataset and the test dataset originate from the same population. In other words, the reference and test datasets are sampled from one population and have a similar data-generating process. This assumption ensures that the pre-trained algorithm based on the reference dataset can be translated into the application on the test dataset. Specifically, our proposed framework focuses on the feature space, and thus a reference dataset with features as diverse as the test dataset will contribute to improving the forecasting performance.

The reference dataset used for training the algorithm is expected to cover the newly given time series in feature spaces, which is a significant concern when we opt for a targeted reference dataset. Recently, \cite{kang2020gratis} propose GRATIS (GeneRAting TIme Series with diverse and controllable characteristics) as an approach to simulating time series based on mixture autoregressive (MAR) models, which provides a guarantee for obtaining sufficient and targeted new time series with controllable features. Instead of simulating time series from models with fixed parameter values as most typical simulation processes do, GRATIS uses parameter distributions to generate time series data based on MAR models, allowing to capture the dependence nature of time series and generate diverse time series instances. Besides, GRATIS can be used to generate sets of time series with target features by tuning the parameters of the MAR models.

\cite{kang2020gratis} design a simulation study to generate $20,000$ yearly, $20,000$ quarterly, and $40,000$ monthly time series with certain parameter settings. They project the features \citep[computed using the R package \texttt{tsfeatures},][]{rtsfeatures} of the generated time series to a two-dimensional instance space and investigate that the features for the generated time series comprehensively cover that for the M4 competition data.

In this paper, we follow \cite{kang2020gratis} and apply the GRATIS approach to separately generate $20,000$ yearly, $20,000$ quarterly, and $40,000$ monthly time series that have the same forecast horizons as the M4 competition data. We use the implementation available in the \texttt{gratis} package for R \citep{rgratis}. The lengths of the generated time series are randomly sampled from the distributions of those of the M4 data (see Figure~\ref{fig:hist}). We refer to the generated time series as the reference dataset. Benefiting from the diversity and coverage of the generated time series in feature spaces, the models trained by the reference dataset in the training phase can be applied to the M4 dataset. The details of the reference and test datasets are summarized in Table~\ref{tab:data}.

\begin{table}[ht]
  \centering
  \caption{The number and forecast horizons of time series in the reference and test datasets.}
  \label{tab:data}
  \resizebox{\linewidth}{!}{
  \begin{tabular}{lccccccc}
  \toprule
  Dataset & & \multicolumn{2}{c}{Yearly} & \multicolumn{2}{c}{Quarterly} & \multicolumn{2}{c}{Monthly} \\
  \cmidrule(lr){3-4} \cmidrule(lr){5-6} \cmidrule(l){7-8}
  \multicolumn{1}{c}{} & & Number & Horizon & Number & Horizon & Number & Horizon \\
  \cmidrule(lr){3-4} \cmidrule(lr){5-6} \cmidrule(l){7-8}
  Reference (GRATIS) & & 20,000 & 6 & 20,000 & 8 & 40,000 & 18 \\
  Test (M4) & & 23,000 & 6 & 24,000 & 8 & 48,000 & 18 \\
  \bottomrule
  \end{tabular}}
\end{table}

\subsection{Time series features}
\label{sec:features}
Time series features contain information that captures the dynamic patterns in data and characterizes their properties as numerical values. There are many time-series analysis methods to depict these characteristics, such as autocorrelation, entropy, statistical tests, and linear and nonlinear regression analysis. The features used in our proposed framework should be able to identify the characteristics of various aspects of the time series.

We consider the set of $42$ features which are the same as the features in \citet{montero2020fforma}. These $42$ features, implemented in the \texttt{tsfeatures} package for R, capture the characteristics of the time series from various aspects. For instance, \textsf{peak} indicates the location of the maximum value in the seasonal component and STL decomposition of the series. The features \textsf{nperiods} and \textsf{seasonal-period} are categorical variables: \textsf{nperiods} takes the values $0$ or $1$, and \textsf{seasonal-period} takes the values $1$, $4$, or $12$ for yearly, quarterly and monthly series, respectively. Multiple dummy variables should be created from the feature \textsf{seasonal-period}: \textsf{seasonal-period-q} (takes the value of $1$ when the value of \textsf{seasonal-period} is $4$ and is otherwise $0$) and \textsf{seasonal-period-m} (takes the value of $1$ when the value of \textsf{seasonal-period} is $12$ and is otherwise $0$). In this way, we actually use $43$ features in our experiment.

\subsection{Interval forecast evaluation}
\label{sec:evaluation}
In this paper, we apply the central $(1- \alpha) \times 100 \%$ PIs for the median to assess the future uncertainty of point forecasts. We use the mean scaled interval score \citep[MSIS,][]{gneiting2007strictly} to measure the accuracy of PIs, as used in the M4 competition. The calculation of MSIS can be stated as follows:
\begin{align}
\label{eq:msis}
\mathrm{MSIS} = \frac{1}{h}\frac{\sum_{t=n+1}^{n+h}(U_t-L_t)+\frac{2}{\alpha}(L_t-Y_t)\mathbbm{1}\left\{ Y_t < L_t\right\} + \frac{2}{\alpha}(Y_t - U_t)\mathbbm{1}\left\{Y_t>U_t\right\}}{\frac{1}{n-m}\sum_{t=m+1}^{n} \vert Y_t-Y_{t-m} \vert},
\end{align}
where $Y_t$ are the true values of the future data, $\left[ L_t, U_t \right]$ are the generated PIs, $h$ is the forecasting horizon, $n$ is the length of the historical data, and $m$ is the time interval symbolizing the length of the time series periodicity (e.g., $m$ takes the values of $1$, $4$, and $12$ for yearly, quarterly, and monthly data, respectively), $\mathbbm{1}$ is the indicator function, which returns $1$ when the condition is true and otherwise returns $0$.

Equation (\ref{eq:msis}) illustrates the logic and calculations of MSIS. The numerator is a penalty for the width of the generated PIs and for the cases where the generated PIs have not covered the true values of the future period. The denominator attempts to make the score less scale dependent.

\section{Feature-based interval forecast combination}
\label{sec:fuma}

\subsection{Linking time series features with interval forecasting performance}
\label{sec:gam}

A crucial step in our proposed time series forecasting framework is to capture how time
series features relate to the interval forecasting performance (MSIS) of each individual
method in a pool. We use generalized additive models (GAMs), which were originally proposed by
\citet{hastie1990generalized}, to characterize the contribution of each time series feature
to the interval forecasting performance in the training phase of our proposed framework,
where the response variable is MSIS and the covariates are the time series features.
Since the values of MSIS are all positive, we take the logarithmic form of the MSIS scores
to expand their range to the real number set $\bm{\mathrm{R}}$. Considering $p$ extracted
features and $M$ forecasting methods, the GAM that we train for the
$j$-th method using $N$ time series in the reference dataset can be written as:
\begin{align}
  \label{eq:gamf}
g(E(\log(\mathrm{MSIS}_{ij}))) = \beta_{j0} + \beta_{j1}F_{1i} + ... + \beta_{jk}F_{ki} + s_{j1}( F_{(k+1)i}) + ... + s_{j(p-k)}(F_{pi}),
\end{align}
where $i=1,...,N$ and $j=1,2,...,M$, $\mathrm{MSIS}_{ij}$ is the MSIS value of the $i$-th time series using the $j$-th method, $\bm{F_{i}}=\left\{F_{1i}, ... , F_{pi} \right\}$ denotes a predictor vector consisting of extracted features of the $i$-th time series, $F_{1i}, ... , F_{ki}$ are linear predictors with dummy features (features that have value as categorical data), $F_{(k+1)i}, ... , F_{pi}$ are predictors that can be modeled non-parametrically except linear terms, $g$ is the link function used to establish the relationship between the mean values of $\log(\mathrm{MSIS})$ and the extracted features,  $\beta_{j0}$ denotes the intercept of the regression, $\beta_{j1}, ... , \beta_{jk} $ are the regression coefficients of the linear terms, and the terms $s_{j1}(\cdot), ... , s_{j(p-k)}(\cdot)$ are smooth and non-parametric functions.

GAMs are flexible but computationally challenging in
determining the form of smooth functions and controlling the smoothness of these
functions. In this paper, we estimate the GAMs by using the penalized iterative least
squares method introduced in the R package \texttt{mgcv} \citep{wood2015package}. By
minimizing the generalized cross-validation score, the method synchronously identifies the
degrees of freedom for each smooth function in the process of model fitting. In GAMs,
the smooth functions in Equation (\ref{eq:gamf}) can be determined by selecting the
appropriate penalty for each pre-prepared basis function, which controls its degrees of
freedom using a single smoothing parameter \citep{wood2001mgcv}.

It is worth mentioning that our framework is general and other nonlinear or nonparametric
approaches are equally well applicable. However, we find that GAM applies to our
situation due to the following key merits:

\begin{enumerate}
\def\labelenumi{\arabic{enumi}.}
\item \textbf{Interpretability}. It is straightforward to explore the partial effects of each
  time series feature on the interval forecasting accuracy. The marginal effect of
  each feature on the MSIS is not interfered by other features due to the additive form of
  the model. The effect analysis, established using GAM, plays a driving role in the design of
  a weight determination mechanism (see Section \ref{sec:threshold} for the details), which
  is dedicated to assigning weights for uncertainty estimation based on the model combining.

\item \textbf{Regularization}. The model is able to prevent over-fitting by controlling
  the smoothness of the predictor functions and adapting a cross-validation scheme. This
  is particularly useful if one has more than necessary features as the covariates.
  Particularly, we consider the set of $43$ features in our experiment.

\item \textbf{Flexibility}. With GAM, smooth functions are no longer restricted to linear
  and polynomial forms, providing excellent performance in capturing the nonlinear
  relationship between time series features and interval forecasting accuracies.
\end{enumerate}

\subsection{Weight assignment and optimal threshold ratio search}
\label{sec:threshold}

Time series forecast combination firstly selects a suitable collection of
forecast models from a model pool and then produces the forecasts
based on their weighted combination. A vast amount of literature shows such a
procedure could produce accurate point forecasting results, especially when none of the
individual models is close to the true model. We extend the idea of forecasting combination
to the prediction interval combination.

We transform the fitted MSIS values of the pre-trained GAMs into a convex set of combining weights to
measure the importance of each individual method in the interval forecasting process with
the adjusted softmax function for the $i$-th time series in the $j$-th individual method
as:
\begin{align}
  \label{eq:adsoftmax}
P_{ij} = \frac{\exp\left\{ \frac{\mu_i-\widehat{\log(\mathrm{MSIS}_{ij})}}{\sigma_i} \right\}}{\sum_{k=1}^{M}\exp\left\{\frac{\mu_i-\widehat{\log(\mathrm{MSIS}_{ik})}}{\sigma_i}\right\}},
\end{align}
where $i=1,...,N$ and $j=1,...,M$, $\mu_i$ and $\sigma_i$ denote the mean and standard
deviation of the fitted values for $\log(\mathrm{MSIS})$ obtained by the $M$ pre-trained GAMs for the $i$-th time series, respectively.

The softmax function normalizes each element in the input vector to a combining weight and
ensures the sum of all the elements in the transformed weight vector is equal to
$1$. The adjusted softmax function is actually a softargmin function, and we normalize
the input elements by $\mu_i$ and $\sigma_i$.  We prefer the exponential form rather than
other forms (e.g., square or absolute form) in the softmax function because negative
values should be down-weighted to near-zero. With the exponential form, a larger
$\log(\mathrm{MSIS})$ value, that is, a poor prediction accuracy, corresponds to a lower
weight compared to others. The adjusted softmax function also avoids a well-known
problem in the original softmax function \citep{goodfellow2016deep} that a larger value of
the input vector leads to a much higher weight than other elements.

We select a subset of appropriate methods for each time series tailored to their features from the method
pool using an optimal threshold ratio searching algorithm.
The pseudocode of the algorithm is presented in Algorithm~\ref{alg:opt}. We
first define a threshold ratio $Tr$ as a random number between $0$ and $1$. For the $i$-th
time series and the $j$-th individual method, we calculate the ratio of weight by
$R_k = P_{ij}/\max(P_{ik})$, where $k = 1,2,\cdots, M$. Then, the individual methods that
satisfy $R_k \geq Tr$ are selected for forecast combination. In particular, $Tr = 0$
indicates that all the methods from the pool are selected, and $Tr = 1$ indicates that only the
method with the minimal fitted $\log(\mathrm{MSIS})$ is selected.
In summary, the algorithm is essentially a searching process that calculates combined forecasts in the configuration of each pre-set threshold ratio, and then determines the threshold ratio with the highest accuracy as the optimal threshold.
Hence, the threshold ratio determines the number of candidate methods selected for model combining.

\begin{algorithm}
  \renewcommand{\algorithmicrequire}{\textbf{Input:}}
  \renewcommand{\algorithmicensure}{\textbf{Output:}}
  \caption{The optimal threshold ratio search}
  \label{alg:opt}
  \begin{algorithmic}
    \Require ~~\\
    $O=\left\{x_1, x_2,..., x_N \right\}$: the collection of $N$ time series in the reference dataset.\\
    $Tr=\left\{Tr_1, Tr_2,..., Tr_q \right\}$: the set of $q$ pre-set threshold ratios.\\
    $M$: the number of individual forecasting methods.
    \Ensure ~~\\
    The optimal threshold ratios for yearly, quarterly and monthly data.
  \end{algorithmic}
  \begin{algorithmic}[1]
    \For{$i=1$ to $q$}
    \For{$j=1$ to $N$}
    \State Obtain the fitted $\log(\mathrm{MSIS})$ of time series $x_j$ from the $M$ pre-trained GAMs in the training phase.
    \State Apply the Equation (\ref{eq:adsoftmax}) to calculate the adjusted softmax transformation $P$ for $x_j$.
    \State Calculate the ratio of $P$: $R_{k}=P_{k}/\smash{\max_{1 \leq k \leq M}} (P_{k})$.
    \State Select the individual methods that satisfy $R_{k} \geq Tr_i$ for $x_j$ and utilize these methods for forecast combination (see Section \ref{sec:comb} for the details).
    \State Calculate the MSIS value of $x_j$.
    \EndFor
    \State Calculate the average MSIS values of yearly, quarterly and monthly data.
    \EndFor
    \State The optimal threshold ratios are pre-set threshold ratios with minimal MSIS for the yearly, quarterly and monthly series in $O$, respectively.
  \end{algorithmic}
\end{algorithm}

\subsection{Prediction interval combination}
\label{sec:comb}

We combine the PIs calculated from the previously selected methods in
Section \ref{sec:threshold}. Inspired by the previous studies on quantiles combination
\citep{hora2004probability,lichtendahl2013better}, we consider two interval combination
methods in this paper, which are the simple average and the weighted average.

The interval combination considers the uncertainty of future forecasts with a
certain set of combining weights. Assuming $S$ forecasting methods are
selected for the $i$-th time series according to a pre-defined threshold ratio, the weighted
lower ($f_{wi}^l$) and upper ($f_{wi}^u$) bounds of the $h$-step prediction interval are defined as:
\begin{align}
\label{eq:interval}
\begin{split}
f_{wi}^l &= \frac{1}{\sum_{k=1}^{S}P_{ik}}\sum_{k=1}^{S}P_{ik}f_{ik}^l, \\
f_{wi}^u &= \frac{1}{\sum_{k=1}^{S}P_{ik}}\sum_{k=1}^{S}P_{ik}f_{ik}^u,
\end{split}
\end{align}
where $f_{ik}^l$ and $f_{ik}^u$ are the lower and upper bounds of the
$h$-step prediction interval for the selected $k$-th individual method, and $P_{ik}$ denotes the weight of the $k$-th method being selected, which is calculated from the adjusted softmax function. If $P_{ik}=1$ for $k=1,2,..., S$, the combined prediction interval
$\left[f_{wi}^l,~ f_{wi}^u \right]$ reduces to the simple average combination.

In addition to PIs, our proposed framework also aims to provide improved point forecasts, giving a comprehensive outlook of the expected future values and the future uncertainty. For $i$-th time series, the $h$-step point forecasts can be calculated as:
\begin{align}
\label{eq:point}
f_{wi} = \frac{1}{2}(f_{wi}^l + f_{wi}^u),
\end{align}
where $f_{wi}$ is the point forecasts for the $i$-th time series. In the same way as the combined prediction intervals, $f_{wi}$ reduces to the simple average combination when $P_{ik}=1$ for $k=1,2,..., S$.

We have developed an R package \texttt{fuma} for the implementation of the aforementioned framework, which is
available at \url{https://github.com/xqnwang/fuma}.

\section{Application to the M4 competition data}
\label{sec:application}

In this section, we apply our approach to the M4 competition data, defining a suitable pool of models that will also act as benchmarks. We also analyze the partial effects of time series features on the interval forecasting accuracy of each individual model. In addition, we present the optimal threshold ratios captured in the reference data, as well as the interval forecasting results of M4 data based on our proposed framework.

\subsection{Evaluation measures}
\label{sec:measure}
To assess the performance of our proposed framework, we consider the MSIS in Equation (\ref{eq:msis}) and the absolute coverage difference (ACD) as the measures of interval forecasting accuracies, as used in the M4 competition. As a supplemental scoring rule, ACD measures the absolute difference between the actual coverage of the method and the nominal coverage, where coverage reflects the rate at which the true values fall within the PIs. Lower MSIS and ACD values are better.

We also evaluate point forecasting accuracies using the mean absolute scaled error \citep[MASE,][]{hyndman2006another}, given by
\begin{align*}
\mathrm{MASE}&=\frac{1}{h}\frac{\sum_{t=n+1}^{n+h}|Y_{t}-\hat{Y}_{t}|}{\frac{1}{n-m}\sum_{t=m+1}^{n}|Y_{t}-Y_{t-m}|},
\end{align*}
where $\hat{Y}_{t}$ are the point forecasts. MASE is considered for its excellent mathematical properties, such as less scale dependent and less insensitive to outliers. Lower MASE values are better.

\subsection{Individual model pool}
\label{sec:setup}
We use eight forecasting models as our method pool, as shown in Table~\ref{tab:methods} and implemented in the R package \texttt{forecast} \citep{hyndman2018forecast}. Note that the forecasting results of sna\"{i}ve models for yearly series essentially coincide with that of na\"{i}ve models and, thus, there are seven forecasting models are considered in the model pool for the yearly series.

\begin{table}
  \centering
  \caption{The model pool considered in the application to the M4 competition data.}
  \label{tab:methods}
  \renewcommand{\arraystretch}{1.2}
  \resizebox{\linewidth}{!}{
  \begin{tabular}{lp{0.8\columnwidth}}
  \toprule
  Individual model & \multicolumn{1}{c}{Description} \\
  \midrule
  auto-arima & The best autoregressive integrated moving average model that is automatically selected by the AICc value. \\
  ets & Exponential smoothing state space model proposed by \citet{hyndman2002state}. \\
  tbats & The exponential smoothing state space model with Trigonometric, Box-Cox transformation, ARMA errors, Trend and Seasonal components. \\
  stlm-ar & Time series is decomposed by STL method proposed by \citet{cleveland1990stl}, then an AR model are fitted for the seasonally adjusted series. \\
  rw-drift & Random walk with drift. \\
  thetaf & A univariate forecasting model proposed by \citet{assimakopoulos2000theta}. It can be seen as a decomposition approach to forecasting by modifying the local curvatures of the time series with Theta-coefficient. \\
  na\"{i}ve & The simplest time series forecasting method. The point forecasts of all forecast horizons are equal to the last observation in the training period. \\
  sna\"{i}ve & Seasonal na\"{i}ve. The point forecast is equal to the most recent value of the same season. \\
  \bottomrule
  \end{tabular}}
\end{table}

\begin{figure}
  \centering \includegraphics[width=1\textwidth]{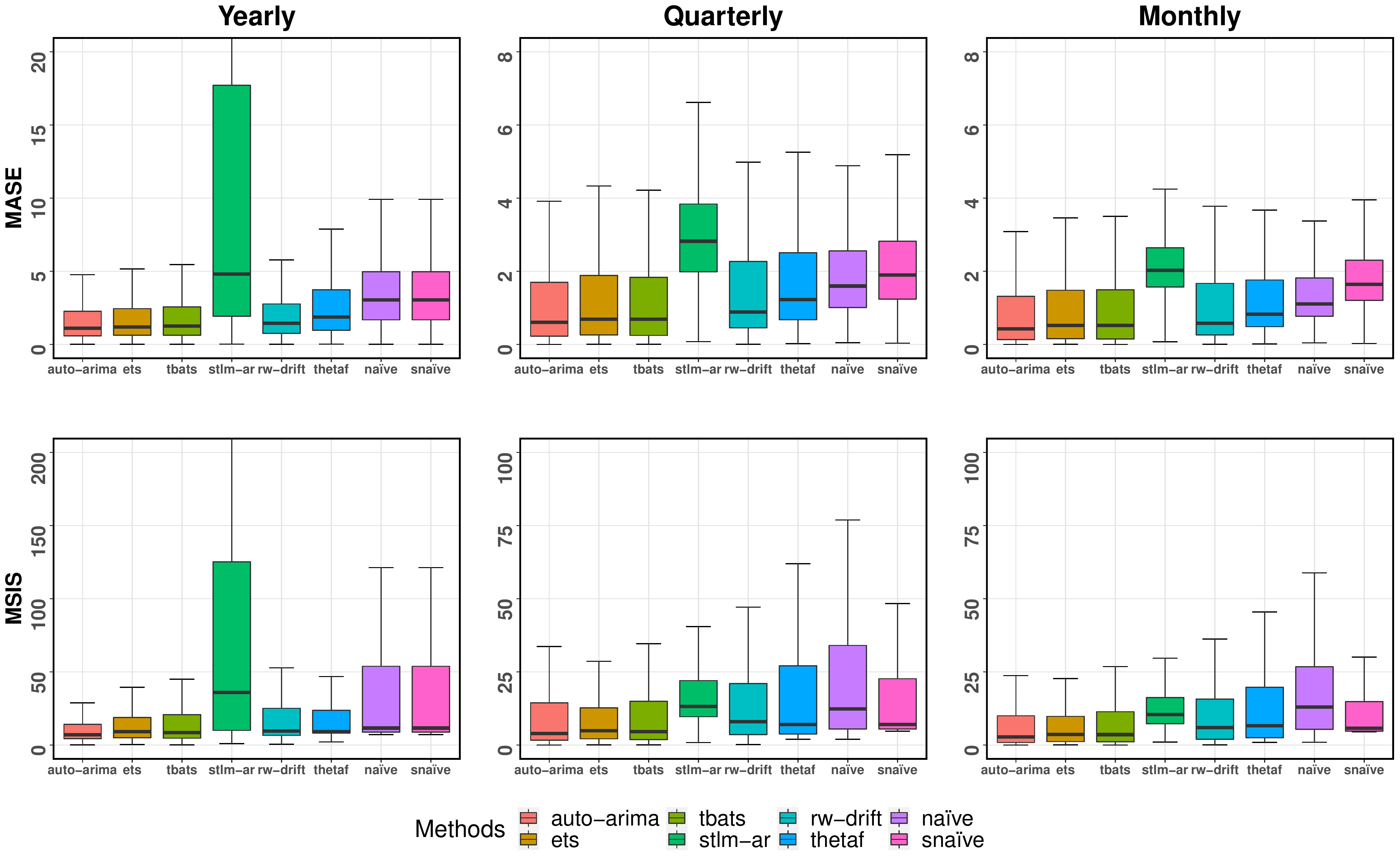}
  \caption{Boxplots of point and $95\%$ interval forecasting accuracy over reference dataset for the individual method pool.}
  \label{fig:boxplot}
\end{figure}

Given the individual model pool, we first calculate the point forecasting accuracy, which is evaluated in terms of MASE and the forecasting accuracy of the $95\%$ confidence intervals ($\alpha = 0.05$), which is measured by MSIS, for all the methods in the pool on the reference dataset. We can see from Figure~\ref{fig:boxplot} that the distributions of point forecasting accuracy for different individual methods are clearly similar to those of the interval forecasting accuracy. For example, for the yearly series, the median and variance of both point and interval forecasting accuracies of the stlm-ar method are significantly larger than that of auto-arima, ets and tbats. Moreover, auto-arima and ets perform well in both point and interval forecasts for yearly, quarterly and monthly series in reference data, while stlm-ar, na\"{i}ve and sna\"{i}ve methods perform poorly compared to other methods in the method pool. This indicates that the proposed forecasting framework for the uncertainty estimation may be used to provide promising point forecasts.

\subsection{Effect analysis of time series features}
\label{sec:effect}
Given the feature matrix $F_{N \times p}$ and score matrix $\mathrm{MSIS}_{N \times M}$ for the $N$ time series in the reference dataset and $M$ individual methods, GAMs are modeled for all the methods in the pool, giving a comprehensive description of the partial effects of features on the interval accuracy of forecasting methods.

\begin{figure}
  \centering \includegraphics[width=1\textwidth]{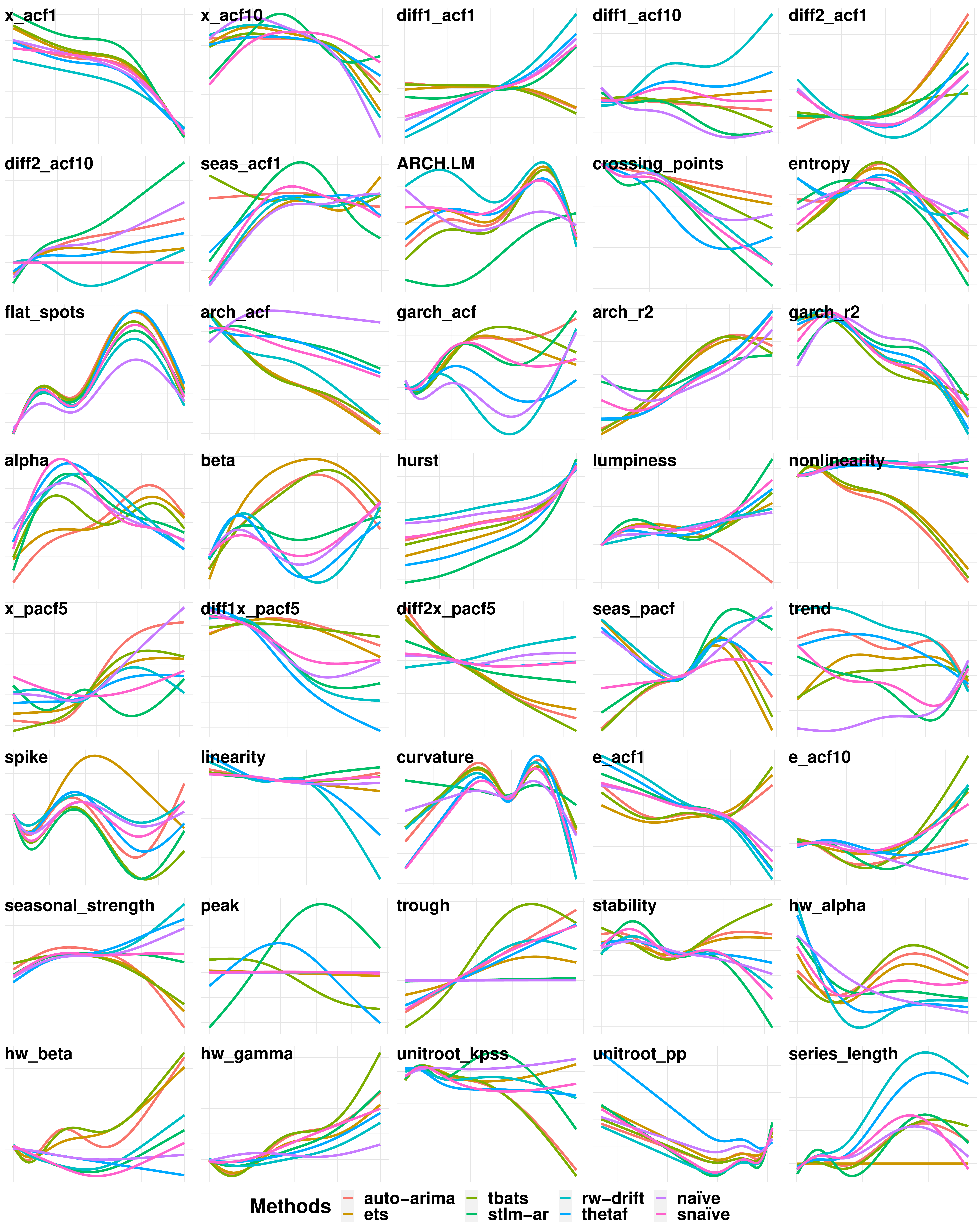}
  \caption{The partial effects of features (x-axis) on $\log(\mathrm{MSIS})$ (y-axis) from trained GAMs for reference data with the method pool. Plots contain $40$ features, and $3$ dummy features are removed.}
  \label{fig:gamplot}
\end{figure}

For demonstration purposes, we combine the partial effect plots of eight individual methods, as presented in Figure~\ref{fig:gamplot}. In the analysis, the MSIS scores assume a $95\%$ confidence level. Note that since the GAMs analysis is performed on the whole reference dataset (including yearly, quarterly and monthly categories generated by GRATIS), the marginal effect of each feature on $\log(\mathrm{MSIS})$ in Figure~\ref{fig:gamplot} is also built based on all the frequencies of the reference dataset and not interfered by other features (e.g., data frequency). We analyze the relationship depicted by GAMs from the following three angles.

Given a particular forecasting method, Figure~\ref{fig:gamplot} first reveals that the partial effect of one feature on the interval forecasting performance is distinct from the other features. This distinction stems from the properties and intrinsic patterns reflected by the various features.
Taking the auto-arima method as an example, if we keep other features fixed, the plot shows a downward trend as the value of \textsf{x-acf1} increases, indicating a drop in the MSIS values, which further implies an improved accuracy.
We consider in detail the cause of this phenomenon: \textsf{x-acf1} reflects the degree of the autocorrelation relationship in the time series, while auto-arima is excellent at capturing the autocorrelation.
As another example, the plot shows an inverted-U shape relationship (a slight rise and then a substantial fall) between \textsf{seasonal-strength}, which measures the seasonal strength, and the MSIS values. The curve indicates that the auto-arima method works well in capturing strong seasonality rather than inconspicuous seasonality of the time series using the seasonal part of the ARIMA model. Therefore, the auto-arima method should be chosen to deal with time series with strong seasonality.

Figure~\ref{fig:gamplot} also indicates that a feature has its unique way of affecting the interval forecasting performance of some specific forecasting methods, while sometimes all the forecasting methods behave similarly with time series features changing.
Taking the feature \textsf{x-acf1} as an example, as the value of \textsf{x-acf1} increases, the interval forecasting performance of the eight methods in the pool successively improves in a similar path under the condition of keeping other features fixed.
In addition, as the value of \textsf{seasonal-strength} increases, the MSIS values of the auto-arima method present an inverted-U shape in a similar way as the ets and tbats methods, while that of other individual methods show an overall ascent. This suggests that we would prefer the auto-arima, ets and tbats methods to others when we have to forecast a time series with a large value of \textsf{seasonal-strength}.
Therefore, the GAMs analysis facilitates the model selection in light of how features affect the interval forecasting accuracy of the methods in a pool.

Finally, Figure~\ref{fig:gamplot} shows that some features are biased towards up-weighting some forecasting methods over others. The time series features, which are applied to select appropriate methods for interval forecasting, should perform discriminatingly on how to affect the forecasting accuracy of various methods. The features with similar growth paths of partial effects on all the individual methods would play a weak role in the model selection process. In contrast, as shown in Figure~\ref{fig:gamplot}, \textsf{diff1-acf1}, \textsf{arch-acf}, \textsf{alpha}, \textsf{beta}, \textsf{lumpiness}, \textsf{non-linearity}, \textsf{seasonal-strength}, \textsf{peak}, \textsf{trough}, and \textsf{hw-beta} may have significant impacts on our model selection process due to their diverse partial effects on the forecasting performance.

\subsection{Interval forecasting results}
\label{sec:results}
\begin{figure}
  \centering \includegraphics[width=1\textwidth]{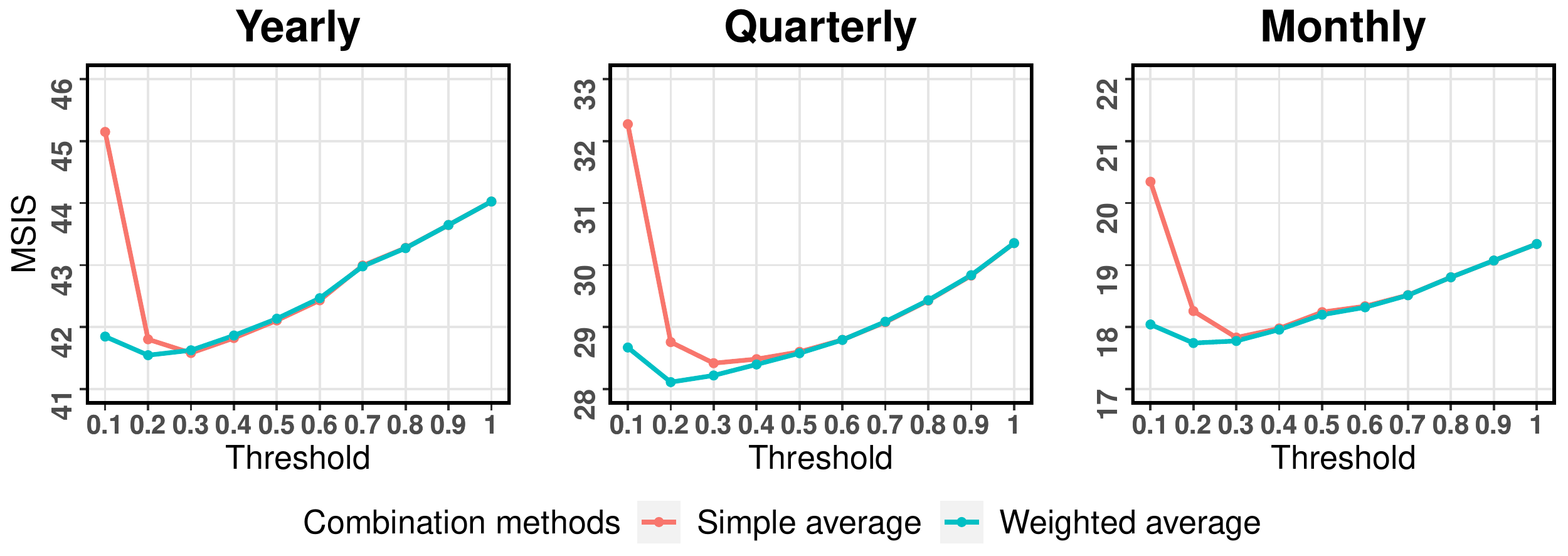}
  \caption{The search path of the optimal threshold ratios for yearly, quarterly and monthly series on the reference dataset. Two interval combination methods are considered: the simple average and the weighted average.}
  \label{fig:threshold}
\end{figure}

We first apply all the pre-trained GAMs to search for the optimal threshold ratios (see Algorithm~\ref{alg:opt} for the details) that performs best on selecting appropriate methods for each data frequency on the reference dataset, visualized in Figure~\ref{fig:threshold}.
A larger threshold value means that fewer methods are selected for model combining, while a smaller threshold value means that many more methods are used for model combining.
As we can see from all the panels, the averaged MSIS scores of each data frequency show an initial decrease and then increase as the threshold increases.
This indicates that controlling the number of methods using the threshold searching algorithm is beneficial for improving the forecasting performance in our experiment.
As presented in Figure~\ref{fig:threshold}, the optimal thresholds for yearly, quarterly and monthly series are all set to $0.3$ and $0.2$ for the simple average and the weighted average combination, respectively.

Having identified the optimal settings for the threshold ratios, given a new time series, we can easily map the optimal thresholds into which models are selected for model combination and what weights are assigned to these models. For example, if the weight values of $0.3,0.3,0.2,0.01,0.06,0.07,0.03$, and $0.03$ are initially assigned to auto-arima, ets, tbats, stlm-ar, rw-drift, thetaf, na\"{i}ve, and sna\"{i}ve using the previously trained GAMs and the adjusted softmax function, respectively, then auto-arima, ets, tbats, rw-drift, and thetaf are selected for model combination because the ratios of their weights ($R_k$) are greater than or equal to the optimal threshold ($0.2$). Subsequently, their weights are normalized to sum to one.

We benchmark the forecasting performance of our proposed feature-based framework, abbreviated from now on simply as `fuma' (forecast uncertainty based on model averaging), against all the methods in the pool as well as their simple equally weighted combinations. We adopt an overall appraisal from the interval forecasting performance as well as the point forecasting performance. All the models considered for comparison are listed as follows:

\begin{itemize}[noitemsep,nolistsep]
\item All the individual models in the model pool. This collection includes eight models that we select in our application on the M4 data, as listed in Table~\ref{tab:methods}.
\item The simple equally weighted combination of all the individual models. We refer to this model as `simple averaging'.
\item The simple combination of individual models selected by the optimal threshold in our proposed framework. We refer to this model as `fuma (mean)'.
\item The weighted combination of individual models that are selected by the optimal threshold in our proposed framework. The weights are determined by the weight assignment mechanism proposed in our framework. We refer to this model as `fuma (weighted)'.
\item The weighted combination of all eight methods in the pool, where the weights are assigned according to our framework. We refer to this model as `fuma (all weighted)'.
\end{itemize}

\begin{figure}
  \centering \includegraphics[width=1\textwidth]{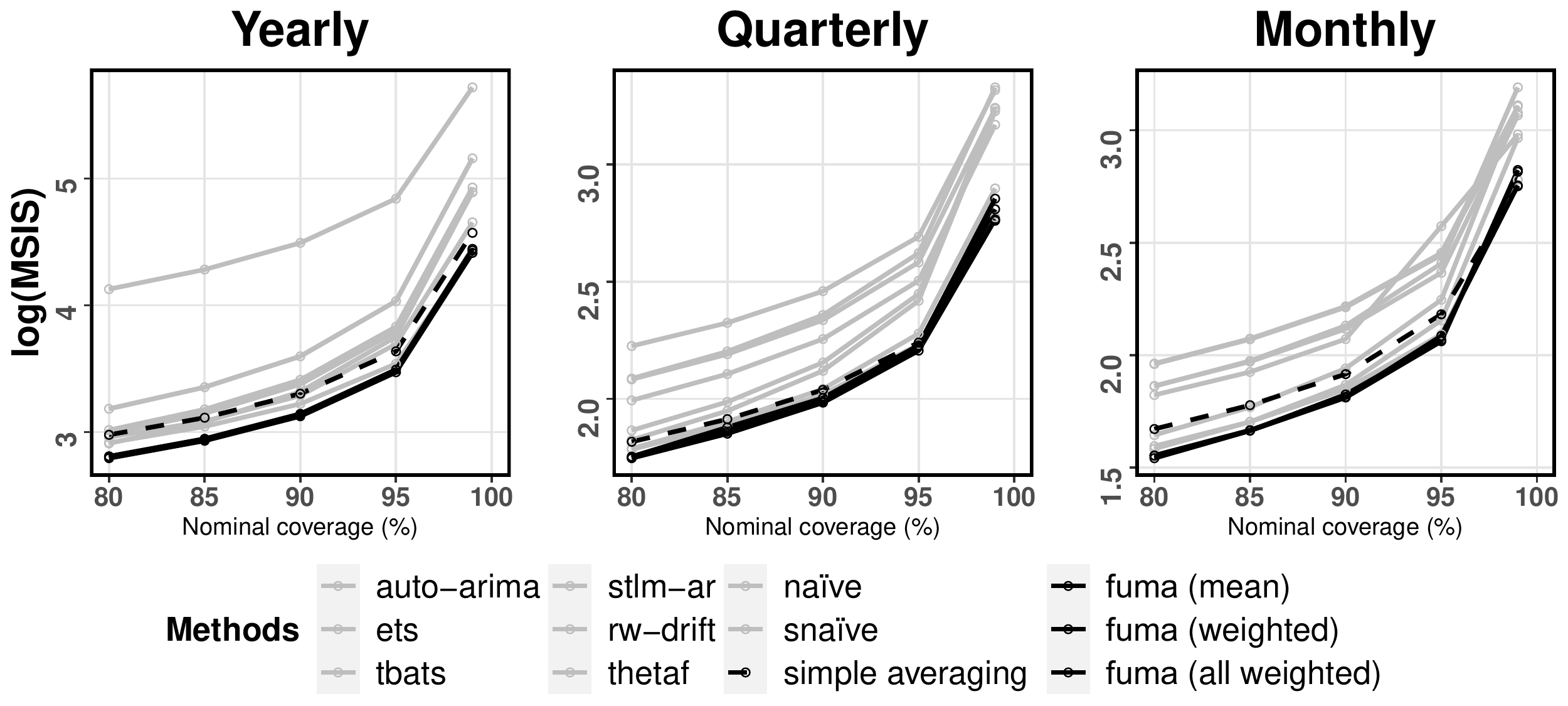}
  \caption{Benchmarking the performance of fuma evaluated in terms of MSIS against the eight individual methods for different confidence levels for each data frequency.}
  \label{fig:performance}
\end{figure}

Figure~\ref{fig:performance} presents the performance of the uncertainty estimation across various confidence levels ($80\%$, $85\%$, $90\%$, $95\%$, and $99\%$) for our feature-based framework and all the benchmark models with regard to the MSIS values. We observe that `fuma (mean)', `fuma (weighted)' and `fuma (all weighted)' consistently outperform all the individual methods as well as `simple averaging' in terms of the MSIS values for each data frequency. The results indicate that the optimal threshold ratio searching algorithm works to select appropriate models for combination, resulting in the fact that `fuma (mean)' achieves performance improvements compared to `simple averaging'. In this way, instead of calculating forecasts of all the models in the pool for the newly given time series, only the individual models selected by the optimal thresholds are expected to be established and serve as the basis for the final forecasts in the testing phase.

\begin{figure}
  \centering \includegraphics[width=1\textwidth]{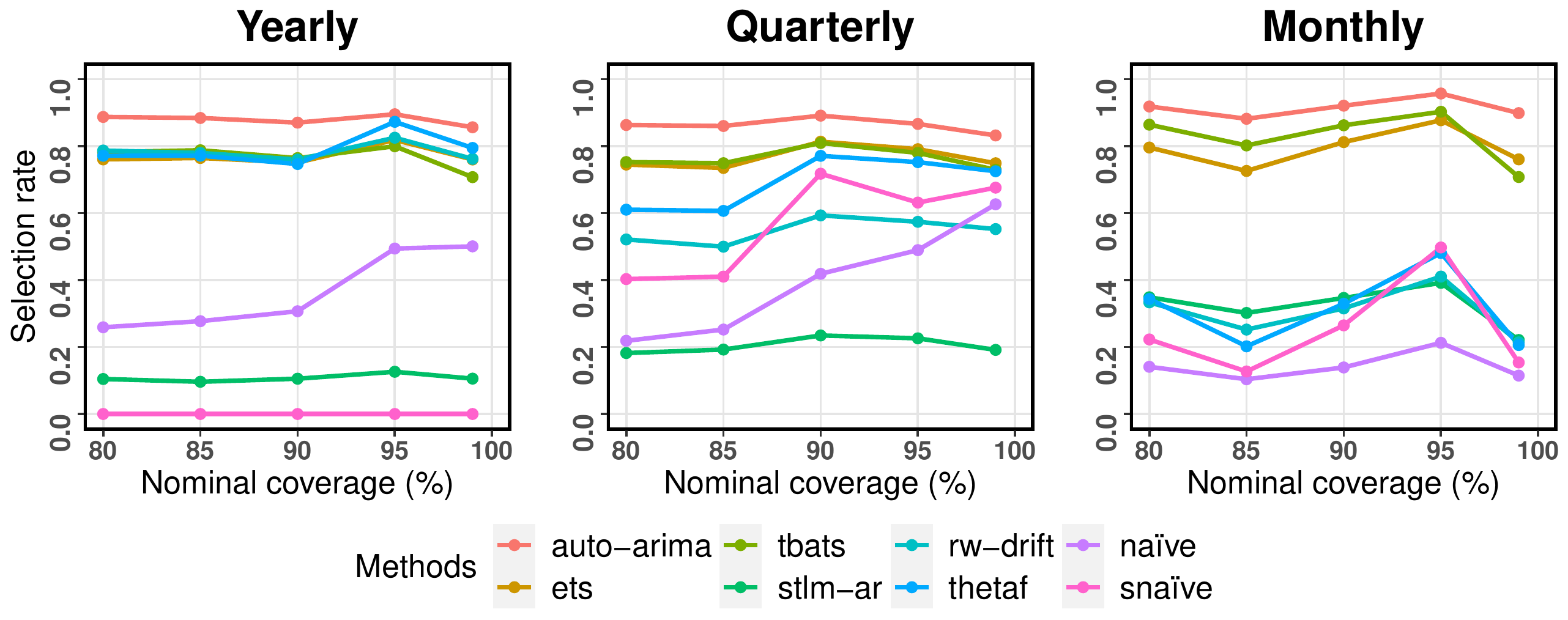}
  \caption{The rates of being selected to combine forecasts in our feature-based framework for each individual method. Different nominal coverages are considered in this plot: $80, 85, 90, 95,$ and $99\%$.}
  \label{fig:select}
\end{figure}

We proceed by comparing the rates of each individual model being selected for `fuma (weighted)' on M4, which are determined by previously trained GAMs and optimal threshold ratios in the testing phase of our framework. Figure~\ref{fig:select} gives a detailed description of the selection rates of each model in the pool for different confidence levels and each data frequency. We can see that auto-arima, tbats and ets are the top three methods that are selected for the weighted combination in our feature-based framework, while the stlm-ar, na\"{i}ve are selected at smaller rates for each data frequency.

We pick two commonly adopted confidence levels ($80\%$ and $95\%$) and summarize the forecasting performance of fuma against all the individual models and their simple average. Table~\ref{tab:results} presents the MSIS and MASE results of all models for each data frequency separately as well as across all frequencies (Total). We observe that the `simple averaging' does not generally help in improving the forecasting performance either for point or interval forecasting. On the other hand, `fuma (mean)', `fuma (weighted)' and `fuma (all weighted)' perform excellently against all the individual methods and their simple average with regards to MSIS and MASE for each data frequency, indicating that fuma gives a comprehensive outlook of the expected future values as well as the future uncertainty.
It is worth mentioning that `fuma (weighted)' produces combined forecasts superior to `fuma (mean)' and `fuma (all weighted)' for monthly data, proving the validity of the weight assignment mechanism and the optimal threshold searching algorithm in our framework.

\begin{table}[ht!]
\caption{Benchmarking the performance of our proposed feature-based framework against all the individual models and their simple equally weighted combination (`simple averaging') with regard to the MSIS and MASE values for the confidence levels of $80\%$ and $95\%$. For each confidence level, the MSIS and MASE values smaller than the minimum value of the model pool and `simple averaging' are marked in bold.}
\label{tab:results}
\centering
\renewcommand{\arraystretch}{1.2}
\resizebox{1\linewidth}{!}{
\begin{tabular}{lrrrrrrrrr}
  \toprule
  & \multicolumn{4}{c}{Confidence level 80\%} & & \multicolumn{4}{c}{Confidence level 95\%} \\
  \cmidrule(lr){2-5}  \cmidrule(lr){7-10}
  & \multicolumn{4}{c}{MSIS} & & \multicolumn{4}{c}{MSIS} \\
  & Yearly & Quarterly & Monthly & Total & & Yearly & Quarterly & Monthly & Total \\
  \cmidrule(lr){2-5}  \cmidrule(lr){7-10}
  auto-arima & 20.450 & 6.224 & 4.901 & 9.000 & & 46.226 & 11.299 & 8.719 & 18.452 \\
  ets & 18.639 & 6.001 & 5.003 & 8.557 & & 34.897 & 9.452 & 8.297 & 15.029 \\
  tbats & 19.291 & 5.981 & 6.192 & 9.310 & & 40.263 & 9.780 & 13.122 & 18.849 \\
  stlm-ar & 62.134 & 9.267 & 6.443 & 20.639 & & 127.747 & 14.805 & 11.140 & 40.297 \\
  rw-drift & 18.433 & 7.471 & 7.420 & 10.099 & & 42.773 & 12.568 & 12.282 & 19.736 \\
  thetaf & 19.826 & 6.480 & 5.209 & 9.069 & & 44.451 & 11.624 & 9.546 & 18.522 \\
  na\"{i}ve & 24.177 & 8.176 & 7.389 & 11.652 & & 56.554 & 14.073 & 12.300 & 23.462 \\
  sna\"{i}ve & 24.177 & 8.071 & 6.502 & 11.178 & & 56.554 & 13.346 & 10.846 & 22.544 \\
  simple averaging & 19.680 & 6.176 & 5.365 & 9.035 & & 38.050 & 9.476 & 9.012 & 16.159 \\
  fuma (mean) & \textbf{16.476} & \textbf{5.772} & \textbf{4.725} & \textbf{7.834} & & \textbf{32.857} & \textbf{9.281} & \textbf{7.900} & \textbf{14.291} \\
  fuma (weighted) & \textbf{16.581} & \textbf{5.749} & \textbf{4.673} & \textbf{7.828} & & \textbf{32.852} & \textbf{9.234} & \textbf{7.859} & \textbf{14.257} \\
  fuma (all weighted) & \textbf{16.336} & \textbf{5.733} & \textbf{4.730} & \textbf{7.793} & & \textbf{32.196} & \textbf{9.075} & \textbf{8.050} & \textbf{14.155} \\
  \cmidrule(lr){2-5}  \cmidrule(lr){7-10}
  & \multicolumn{4}{c}{MASE} & & \multicolumn{4}{c}{MASE} \\
  & Yearly & Quarterly & Monthly & Total & & Yearly & Quarterly & Monthly & Total \\
  \cmidrule(lr){2-5}  \cmidrule(lr){7-10}
  auto-arima & 3.451 & 1.175 & 0.926 & 1.600 & & 3.451 & 1.175 & 0.926 & 1.600 \\
  ets & 3.444 & 1.161 & 0.948 & 1.606 & & 3.444 & 1.161 & 0.948 & 1.606  \\
  tbats & 3.437 & 1.186 & 1.053 & 1.664 & & 3.437 & 1.186 & 1.053 & 1.664 \\
  stlm-ar & 10.387 & 2.028 & 1.334 & 3.701 & & 10.387 & 2.028 & 1.334 & 3.701 \\
  rw-drift & 3.068 & 1.330 & 1.180 & 1.675 & & 3.068 & 1.330 & 1.180 & 1.675 \\
  thetaf & 3.375 & 1.231 & 0.970 & 1.618 & & 3.375 & 1.231 & 0.970 & 1.618 \\
  na\"{i}ve & 3.974 & 1.477 & 1.205 & 1.944 & & 3.974 & 1.477 & 1.205 & 1.944 \\
  sna\"{i}ve & 3.974 & 1.602 & 1.260 & 2.003 & & 3.974 & 1.602 & 1.260 & 2.003 \\
  simple averaging & 3.691 & 1.243 & 0.981 & 1.703 & & 3.691 & 1.243 & 0.981 & 1.703 \\
  fuma (mean) & \textbf{3.031} & \textbf{1.144} & \textbf{0.913} & \textbf{1.484} & & \textbf{3.049} & \textbf{1.147} & \textbf{0.906} & \textbf{1.486} \\
  fuma (weighted) & \textbf{3.037} & \textbf{1.141} & \textbf{0.905} & \textbf{1.481} & & \textbf{3.032} & \textbf{1.142} & \textbf{0.902} & \textbf{1.478} \\
  fuma (all weighted) & \textbf{3.016} & \textbf{1.144} & \textbf{0.910} & \textbf{1.479} & & \textbf{3.023} & \textbf{1.145} & \textbf{0.912} & \textbf{1.482} \\
  \bottomrule
\end{tabular}}
\end{table}

\begin{figure}
  \centering \includegraphics[width=1\textwidth]{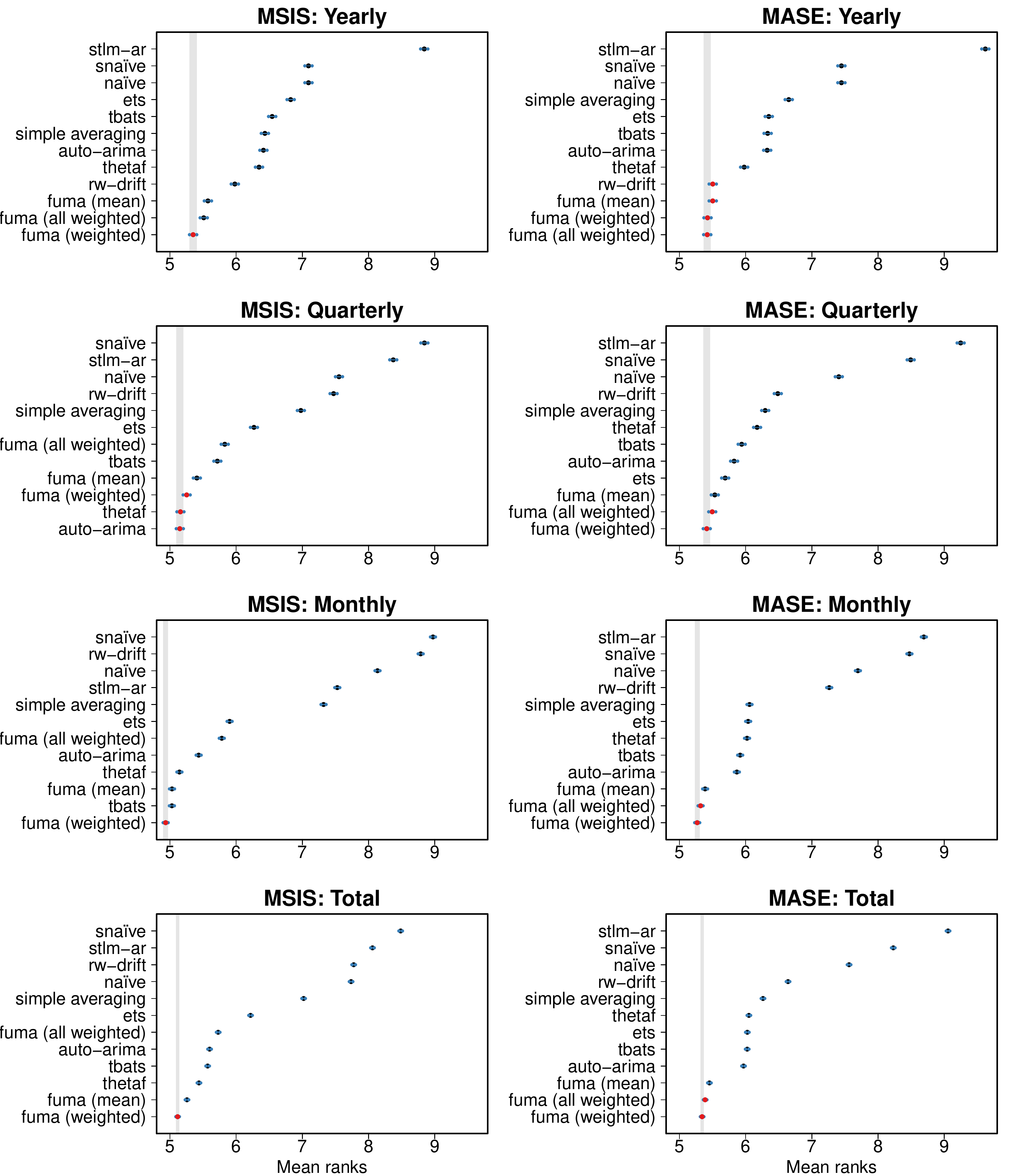}
  \caption{MCB test results for the ranks of all models (individual model pool, `simple averaging', and fuma) for each data frequency separately as well as across all frequencies (Total). The MSIS and MASE values assume a $95\%$ confidence level.}
  \label{fig:nemenyi_95_all}
\end{figure}

Next, we investigate the statistical significance of the performance improvements achieved by fuma. We conduct the multiple comparisons with the best \citep[MCB:][]{koning2005m3} test to identify whether the average ranking differences of all models considered for comparison across time series are statistically significant. The MCB test is applied based on MSIS and MASE for the $95\%$ confidence level, as shown in Figure~\ref{fig:nemenyi_95_all}. With MCB, the ranking performances are statistically different if the intervals of two models do not overlap.

The MCB results show that `fuma (weighted)' results in the best-ranked performance in terms of the MSIS values, except that it ranks similarly with the auto-arima and thetaf methods on the quarterly series with the $95\%$ confidence level. In particular, the interval forecasting performance of `fuma (weighted)' is significantly better than that of both `fuma (mean)' and `fuma (all weighted)' for each data frequency separately as well as across all frequencies, which further confirms the positive effects of the weight assignment mechanism and the optimal threshold searching algorithm. Besides, even if our proposed framework is proposed for interval forecasting, fuma provides comparable and even significantly better point forecasting performance.

Table~\ref{tab:m4rank} depicts the MSIS and ACD results assuming a $95\%$ confidence level for fuma and the top five ranked methods from the M4 competition in terms of PIs precision. We observe that fuma results in comparable performances with the top ranked methods from the M4 competition and `fuma (all weighted)' ranks third for both MSIS and ACD. Specifically, the proposed fuma method ranks second for quarterly and monthly series with regard to MSIS. However, we should treat these comparisons with care, as the participants in the M4 competition did not have access to the test data.

\begin{table}
\caption{Benchmarking the performance of our proposed framework against the top five methods in the M4 competition in terms of PIs precision.}
\label{tab:m4rank}
\centering
\renewcommand{\arraystretch}{1.2}
\resizebox{1\linewidth}{!}{
\begin{tabular}{lrrrrrrrrr}
  \toprule
  & \multicolumn{4}{c}{MSIS} & & \multicolumn{4}{c}{ACD} \\
  \cmidrule(lr){2-5} \cmidrule(lr){7-10}
  & Yearly & Quarterly & Monthly & Total & & Yearly & Quarterly & Monthly & Total \\
  \midrule
  Rank & \multicolumn{9}{c}{M4 competition}\\
  1 (Smyl) & 23.898 & 8.551 & 7.205 & 11.587 & & 0.003 & 0.004 & 0.005 & 0.004\\
  2 (Montero-Manso, et al.) & 27.477 & 9.384 & 8.656 & 13.397& & 0.014 & 0.016 & 0.016 & 0.016 \\
  3 (Doornik, et al.) & 30.200 & 9.848 & 9.494 & 14.596 & & 0.037 & 0.029 & 0.054 & 0.044 \\
  4 (ETS - Standard for comparison) & 34.900 & 9.452 & 8.297 & 15.030 & & 0.111 & 0.018 & 0.016 & 0.040 \\
  5 (Fiorucci \& Louzada) & 35.844 & 9.420 & 8.029 & 15.115 & & 0.164 & 0.056 & 0.028 & 0.068 \\
  \cmidrule(lr){2-10}
  Method & \multicolumn{9}{c}{Our framework}\\
  fuma (mean) & 32.857 & 9.281 & 7.900 & 14.291 & & 0.124 & 0.037 & 0.018 & 0.048 \\
  fuma (weighted) & 32.852 & 9.234 & 7.859 & 14.257 & & 0.128 & 0.036 & 0.016 & 0.048 \\
  fuma (all weighted) & 32.196 & 9.075 & 8.050 & 14.155 & & 0.115 & 0.027 & 0.005 & 0.037 \\
  \bottomrule
\end{tabular}}
\end{table}

\section{Conclusions}
\label{sec:conclusion}
In this paper, we focused on the uncertainty estimation of feature-based time series forecasts where the interest is in forecasting large collections of time series. To this end, we designed a general feature-based time series forecasting framework to explore how time series features affect the uncertainty estimation of forecasts and then translated these findings into an attempt to improve the forecasting accuracy of PIs. At the same time, we developed a new weight determination mechanism, which is applied to assign combination weights for each time series tailored to their features, and an optimal threshold ratio searching algorithm, which focus on selecting the subset models for model combining. To our knowledge, this is the first time that features are taken into account to estimate the uncertainty of forecasts. We investigated the performance of our approach against the benchmark models and the top ranked methods from the M4 competition. We found that our approach performs excellently against the individual benchmark models. Moreover, we demonstrated the positive role of the weight assignment mechanism and the optimal threshold searching algorithm in improving forecasting performance.

\section*{Acknowledgements}
Yanfei Kang is supported by the National Natural Science Foundation of China (No.11701022)
and the National Key Research and Development Program (No. 2019YFB1404600). Feng Li is
supported by the National Natural Science Foundation of China (No. 11501587) and the
Beijing Universities Advanced Disciplines Initiative (No. GJJ2019163).
Petropoulos' work was completed during his visit at the Beihang University in April-May 2019. This research was
supported by the high-performance computing (HPC) resources at Beihang University.

\bibliographystyle{apacite}
\bibliography{myreference.bib}


\end{document}